\documentclass[11pt,twoside]{article}

\newcommand{\wt}{\widetilde}
\newcommand{\ve}{\varepsilon}

\usepackage{geometry}
\usepackage{fancyhdr}
\usepackage{amssymb}
\usepackage{mathrsfs}
\usepackage{mathtools}
\usepackage{centernot}
\usepackage{soul}
\usepackage{xcolor}

\usepackage[export]{adjustbox}
\usepackage{tikz}
\usepackage{caption}
\usepackage{subcaption}

\numberwithin{equation}{section}

\usepackage[affil-it]{authblk}

\usepackage{abstract}

\usepackage{titlesec}
\titleformat*{\section}{\large\bfseries}

\usepackage[labelfont={bf},textfont={small},width=230pt]{caption}

\usepackage[sorting=none, citestyle=numeric-comp, bibstyle=ieee]{biblatex}
\usepackage[colorlinks=true, linkcolor=blue, urlcolor=blue, citecolor=blue,]{hyperref}

\newcommand{\articletitle}{
	Solutions of Koopman-von Neumann equations,\\ their superpositions, and uncertainties
}

\newcommand{\shorttitle}{
	Solutions of KvN equations, their superpositions, and uncertainties
}

\title{
	\Large{
		\articletitle
	}
}

\author{
	Mustafa Amin
	\thanks{\href{mailto:m.amin@uleth.ca}{m.amin@uleth.ca}}
}

\author{
	Mark A. Walton
	\thanks{\href{mailto:walton@uleth.ca}{walton@uleth.ca}}
}

\affil{ \small
	Department of Physics \& Astronomy, University of Lethbridge, Lethbridge, AB, Canada, T1K 3M4
}

\date{}

\begin{document}

\maketitle

\begin{abstract}
	The so-called Koopman-von Neumann (KvN) formulation brings classical mechanics to Hilbert space,
	but many techniques familiar from quantum mechanics remain missing.
	We use the gauge freedom of the KvN equation to derive separable solutions of the KvN equation,
	and by extension the classical Liouville equation.
	The canonical ensemble solution of statistical mechanics is shown to arise from separable solutions.
	We discuss the Hermiticity of the Liouvillian operator,
	and treat a case where its spectrum is discrete.
	Finally,
	we analyze an uncertainty relation between the so-called dynamical time and the Liouvillian,
	and argue that such a relation is natural and can be understood entirely within a classical framework.


\end{abstract}

\section{Introduction}\label{sec:intro}

The Hilbert-space formulation of physical theories has evolved in close connection with quantum theory.  The association of quantum mechanics with Hilbert space is so strong that quantum mechanics is sometimes said to ``live'' in Hilbert space, while classical mechanics does not.  This belief is often used to explain the difference between quantum and classical physics.

Of course,
physical theories do not ``live'' in one mathematical formulation or the other.
As early as 1931,
Koopman
and later von Neumann in 1932,
showed that classical mechanics can be treated in Hilbert space~\cite{Koopman1931, neumann_zur_1932, neumann_zusatze_1932}.
On the other hand,
quantum mechanics has been formulated in the traditionally classical language of phase space,
see~\cite{Hancock2004, Zachos2005, Case2008, Curtright2014}.
Aside from the mathematical formulation used,
even certain physical concepts usually believed to belong exclusively to the quantum realm,
like noncommutativity,
interference,
and other surprising phenomena have been shown to arise in classical theories~\cite{spekkens_defense_2007, bartlett_reconstruction_2012, spekkens_quasi-quantization_2016, catani_why_2023, catani_what_2022}.

Our focus here will be on the Hilbert-space approach to classical mechanics.
	This approach came to be known as the Koopman-von Neumann (KvN) formulation of classical mechanics.\footnote{
	During a late stage of the preparation of this manuscript,
	the authors learned of a recent study of the history of development of the formulation,
	presented in Ref.~\cite{barandes_history_2026}.
	It argues that the Hilbert-space formulation we use here was pioneered instead by Mario Schönberg in 1952 and 1953,
	with key contributions from
	Angelo Loinger,
	Giacomo Della Riccia,
	Norbert Wiener,
	and E. C. G. Sudarshan.
}
There, the state is encoded in a probability amplitude whose absolute square is a probability distribution obeying the classical Liouville equation of motion.  The probability amplitude itself obeys a Schrödinger-like equation, called the KvN equation.  Solving it is tantamount to solving the Liouville equation, but with the advantage of employing the tools of Hilbert space.  Noncommutative variables, such as the Liouvillian operator, arise naturally in this theory, along with their associated uncertainty relations.  With mathematical objects as such, the Koopman-von Neumann theory might be useful for studying the foundational commonalities with, and distinctions from, quantum physics.  Beyond foundations, combining classical and quantum mechanics into a single framework has practical benefits for simplifying the study of some systems by treating one part as classical and the other quantum.  A common language is a good step towards that.

However, even though KvN classical mechanics involves probability amplitudes and noncommutative variables, an understanding of these concepts in a classical setting is still lacking.  Moreover, it is possible that some technical aspects of the formulation may be too unwieldy to allow for the comfortable use of its features.  Here, we attempt to address some aspects of these issues.  Our goal is to show that Hilbert-space techniques are useful for classical statistical mechanics, and to show that some typically quantum phenomena, like uncertainty relations, can be analyzed from a purely classical point of view.

First, we show that using a gauge freedom inherent in the KvN formulation enables the technique of separation of variables for solving the equation of motion in Sec.~\ref{sec:sepvar}.  As mentioned above, and detailed in Sec.~\ref{sec:prob-amp}, solutions to the KvN equation translate into solutions to the Liouville equation, which cannot be solved by separation of variables even for simple Hamiltonians.  It is shown that the canonical ensemble solution results as a separable solution.  Conceptually, the method of separation of variables produces probability distributions lacking correlations between the separated variables.  Further, due to the linearity of the KvN equation, separable solutions can be superposed to produce distributions with correlations between variables.  In the same way, time-independent (equilibrium) solutions can be superposed to produce time-dependent ones.

Second,
in order to allow for the superposition of solutions to differently-gauged KvN equations,
we detail a method of bringing these solutions to the same gauge,
and show the conditions necessary in Sec.~\ref{sec:superposition}.
This ``cross-gauge superposition'' is needed since the separable solutions mentioned above require different gauges.
With the ability to construct superposable probability amplitudes,
we turn to the construction of orthonormal eigenstates of the Liouvillian operator in Sec.~\ref{sec:orthohermit}.
There,
we discuss the Hermiticity of the Liouvillian,
and use a canonical transformation to phase-space coordinates comprised of the Hamiltonian and its Poisson canonical conjugate,
the so-called dynamical time~\cite{bhamathi_time_2003},
to show a formal method of constructing such orthonormal eigenstates.
We also show that in certain cases,
such as that of the simple harmonic oscillator,
the eigenvalues of the Liouvillian operator are discrete.
In Sec.~\ref{sec:illust},
the classical simple harmonic oscillator is analyzed,
and the spectrum of the Liouvillian is found to be similar,
but not identical,
to that of the quantum Hamiltonian.

Finally, we attempt to address some conceptual aspects of the KvN formulation.  Specifically, we address the appearance of noncommuting operators in it.  The presence of these operators in a classical theory, it seems, is not fully understood.  In Sec.~\ref{sec:LUC}, we use one example to show that, while classical noncommutativity may seem surprising, the resulting uncertainty relation is at once similar to quantum mechanical uncertainty and can be entirely understood in classical terms.  In the same vein, we discuss the ``collapse'' of a classical probability amplitude.  This analysis places quantum mechanical concepts in comparison with classical ones, hoping to help better understand them in both theories.

This paper is arranged as follows.
In Sec.~\ref{sec:prob-amp} we briefly introduce the KvN formulation and the gauge freedom associated with its equation of motion.
We use that freedom to introduce the method of separation of variables to solve the equation of motion in Sec.~\ref{sec:sepvar} where we derive the canonical ensemble distribution from the separability of the equation.
Section~\ref{sec:superposition} contains the method for combining solutions to differently-gauged KvN equations to construct another valid solution to the Liouville equation.
In Sec.~\ref{sec:orthohermit} we discuss the Hermiticity of the (gauged) Liouvillian operator and find its eigenfunctions.
An example of the simple harmonic oscillator is presented in Sec.~\ref{sec:illust}.
Finally,
Sec.~\ref{sec:LUC} contains a discussion of the extra variables of KvN which do not commute with the phase-space ones, their uncertainty relations, and state collapse.

\section{The phase-space probability amplitude and its equations}\label{sec:prob-amp}

The time-evolution of classical systems can be described using Hamilton's equations, or equivalently, using the Liouville equation
\begin{align}\label{eq:Liouville}
	\partial_t \rho + \{\rho, H\} = 0~.
\end{align}
Here $\partial_x$ denotes a partial derivative with respect to $x$, $H$ is the Hamiltonian, $(q, p)$ are Poisson canonical conjugates (like position and momentum), and the Poisson bracket $\{\cdot\,,\cdot\}$ is defined by
\begin{align}
	\{f, g\} := \partial_q f \, \partial_p g - \partial_q g \, \partial_p f~.
\end{align}
The Liouville equation describes the evolution of a phase-space probability distribution $\rho(q, p, t)$ for classical systems.  If one has complete knowledge of the system, then said probability distribution is a delta function on phase-space.  The two equivalent descriptions of classical mechanics (Hamilton's equations and the Liouville equation) are akin to the Heisenberg and Schrödinger pictures of quantum mechanics~\cite{Sudarshan2016}.  The ``Schrödinger picture'' of classical mechanics lends itself to situations where complete knowledge is not available; the Liouville equation is central to statistical mechanics.

Unlike the Schrödinger equation, which evolves a complex \textit{probability amplitude}, the Liouville equation describes the evolution of a real probability density.
Usually, the complex probability amplitude itself obeys the Liouville equation~\eqref{eq:Liouville}.  See, for example, \cite{Mauro2002, Gozzi2004}.

It is easy to see, however, that the equation for the classical phase-space probability amplitude is not unique.  Let $\chi(q, p, t)$ be the complex probability amplitude, such that
\begin{align}\label{eq:chi}
	|\chi|^2 = \rho~.
\end{align}
Inserting that into the Liouville equation~\eqref{eq:Liouville} we get
\begin{align}
	\frac{1}{\chi^*} \big[ \partial_t \chi^* + \{\chi^*, H\} \big]
	= - \frac{1}{\chi} \big[ \partial_t \chi + \{\chi, H\} \big]~.
\end{align}
Note that this is a consequence of the Liouville equation containing only first-order derivatives.  We see that $\frac{1}{\chi} \big[ \partial_t \chi + \{\chi, H\} \big]$ can be any imaginary quantity (if $H$ is real), and we can write
\begin{align}\label{eq:gaugedKvN}
	\partial_t \chi + \{\chi, H\} = \frac{1}{i\hbar} \alpha \, \chi
\end{align}
where $\alpha(q, p, t)$ is an arbitrary real function with the dimension of energy.  Similar equations (in a different form) have appeared in~\cite{Bondar2012, Klein2017}.  We chose to divide $\alpha$ by a constant with the dimension of action to give $\wt{H}$ (see below) the dimension of energy.  Since $\alpha$ is arbitrary, the numerical value of the constant is irrelevant, we chose it to be $\hbar$ to facilitate comparison with analogous quantum mechanical expressions; no physical significance is attached to this choice.  The inclusion of $\hbar$ in these formulas is sometimes used in the KvN literature, see Refs.~\cite{Klein2017, Bondar2019} for example.  For any choice of $\alpha$, the Liouville equation for $|\chi|$ is satisfied.

Eq.~\eqref{eq:gaugedKvN} can be rewritten in a way that is formally similar to the Schrödinger equation as
\begin{align}\label{eq:SchKvN}
	i\hbar \, \partial_t \chi = \wt{H} \chi
\end{align}
where the ``tilde-Hamiltonian'' $\wt{H}$ is an operator defined as
\begin{align}\label{eq:tildeH}
	\wt{H} = i\hbar \{H, \cdot\} + \alpha(q, p, t)
\end{align}
and plays a similar role to the Hamiltonian operator of quantum mechanics; namely, it is the generator of time translations.  For the case when $\alpha$ is chosen to be zero, $\wt{H}$ is sometimes called the Liouvillian.  We sometimes refer to the tilde-Hamiltonian, including the arbitrary function $\alpha(q, p, t)$, as the gauged Liouvillian.

In Liouvillian mechanics, the Hamiltonian $H$ defines the physics, and the Liouville distribution $\rho$ defines the state.  The generalized KvN formulation has the tilde-Hamiltonian~\eqref{eq:tildeH} for the physics and $\chi$ for a state.  Both $\wt{H}$ and $\chi$ have redundancies: $\alpha$ for the former and the phase for the latter.  This redundancy is instrumental for some techniques we will employ in later sections, and is reminiscent of gauge freedom.  Observe that the general KvN equation~\eqref{eq:SchKvN}
\begin{align}
	i \hbar \, \partial_t \chi = \wt{H} \chi = i\hbar \{H, \chi\} + \alpha \chi 
\end{align}
remains invariant under the transformation
\begin{align}\label{eq:transformation}
	\begin{split}
	\chi \to \chi' = \chi e^{\frac{i}{\hbar} \varphi}~, \quad
	\alpha \to \alpha' = \alpha - D \varphi~,
	\end{split}
\end{align}
with $D := \partial_t - \{H,\cdot\}$.\footnote{
	We may write Eq.~\eqref{eq:SchKvN} in terms of a covariant derivation $\mathscr{D} := D - \frac{1}{i\hbar} \alpha$ as $\mathscr{D} \chi = 0$.  It is covariant because $\mathscr{D} \chi \to \mathscr{D}' \chi' = (\mathscr{D} \chi) e^{\frac{i}{\hbar} \varphi}$.
}

Equation~\eqref{eq:gaugedKvN} (or~\eqref{eq:SchKvN}) then is a class of equations for the probability amplitude $\chi$, all of which are equivalent to one and the same Liouville equation for the probability density $\rho$.  We call them gauged KvN equations since the choice of $\alpha$ acts as a gauge: it does not alter the physical description of the system.\footnote{
	While the gauge field of electromagnetism is related directly to the physically real electromagnetic field, $\alpha$ in the above equations is one step further from reality.  The function $\alpha$ relates to the phase of the probability amplitude $\chi$, which is in turn related to the physically relevant probability distribution.  We use the term ``gauge'' following some KvN literature.
}
The gauge freedom/nonuniqueness of the dynamical equation for the phase-space probability amplitude is well known, see~\cite{Klein2017, Bondar2019} for example.  Indeed, in~\cite{Bondar2019} the function $\alpha$ appears in the form $\Phi(q,p) + A_p(q,p) \partial_q H - A_q(q,p) \partial_p H$, where $(\Phi, A_q, A_p)$ is a $U(1)$ gauge potential.  In these examples, however, a particular choice of $\alpha$ is emphasized as most suitable for the purpose of the study.  By contrast, here, the full gauge freedom is used to construct $(q,p)$-separable eigenstates of the tilde-Hamiltonian/Liouvillian.

\section{Separable solutions and the canonical ensemble}\label{sec:sepvar}

This section illustrates the utility of the $\alpha$ gauge freedom in finding simple solutions of the Liouville equation.  Consider a time-independent Hamiltonian.  If we choose $\alpha$ to be time-independent then so becomes the tilde-Hamiltonian operator.  Now time is separable in the gauged KvN equation~\eqref{eq:SchKvN} and we can look for solutions of the form
\begin{align}\label{eq:stationary}
	\chi_\ve(q,p,t) = F_\ve(q,p) e^{-\frac{i}{\hbar} \ve t}~,
\end{align}
with $\chi_\ve$ obeying the time-independent equation
\begin{align}\label{eq:timeind}
	\wt{H} \chi_\ve = \ve \chi_\ve~.
\end{align}
The separation constant $\ve$ is an eigenvalue of $\wt{H}$, it has the dimension of energy, and is real if $\wt{H}$ is Hermitian.  More on the Hermiticity of $\wt{H}$ is presented in Sec.~\ref{sec:orthohermit}.

It is evident that the eigenfunctions $\chi_\ve$ of $\wt{H}$ are stationary/equilibrium states: upon absolute-squaring them we obtain probability distributions with no explicit time-dependence.  Up to this point, $\rho = |\chi_\ve|^2$ can be any (normalizable) function of the Hamiltonian and possibly other constants of motion, but the form of the equilibrium state is yet to be specified.  In statistical mechanics, physical and statistical arguments are used to determine which equilibrium ensemble is to be used for a given situation.  One may ask, if separating $t$ produced solutions with the physical interpretation of equilibrium, what equilibrium state would separability in $q$ and $p$ produce?

A separable solution $\chi_\ve(q,p)$ is a product state in $q$ and $p$ of the form $\chi_\ve(q,p) = \chi_\ve^q(q) \chi_\ve^p(p)$.  This produces a joint probability distribution $\rho(q,p) = |\chi_\ve(q,p)|^2 = |\chi_\ve^q(q)|^2 \cdot |\chi_\ve^p(p)|^2$ that is the product of the marginal probability distributions for $q$ and $p$.  Thus, separable solutions describe states where $q$ and $p$ are independent, there is no correlation between them: knowledge gained about one does not yield information about the other.  Superposition between the product states as derived here can be used to make distributions that correlate the canonical variables.  As shown below, these $(q,p)$-separable solutions produce the canonical ensemble.

Consider single-particle Hamiltonians that are additive functions of $q$ and $p$
\begin{align}
	H(q, p) = K(p) + U(q)~.
\end{align}
Note that the variables $q$ and $p$ can be any canonical phase-space coordinates (i.e., $\{q, p\} = 1$) for the system that puts the Hamiltonian in the separable form above, they do not necessarily have to be position and momentum.  With this Hamiltonian, the time-independent equation~\eqref{eq:timeind} becomes
\begin{align}\label{eq:qpalmost}
	\frac{1}{\chi_\ve} \left[
		\frac{1}{U'(q)} \partial_q \chi_\ve
		- \frac{1}{K'(p)} \partial_p \chi_\ve
		\right] = \frac{i}{\hbar} \frac{\ve - \alpha(q,p)}{U'(q) K'(p)}
\end{align}
where a prime denotes a derivative with respect to the argument.  (Note the arbitrary function $\alpha(q,p)$ on the right-hand-side.)  Dividing by $U'(q) K'(p)$ excludes the free particle from this derivation.  Nonetheless, the end result will still be valid for these systems, these are trivially separable anyway.

For Eq.~\eqref{eq:qpalmost} to admit $(q,p)$-separable solutions $\chi_\ve = \chi^q_\ve(q) \chi^p_\ve(p)$, the right-hand-side of the equation must be the sum of functions of $q$ and functions of $p$, say $\frac{i}{\hbar}\left[f_\ve(q) - g_\ve(p)\right]$.  Here $f_\ve(q)$ and $g_\ve(p)$ are arbitrary real functions.  Then the equation becomes separable in $q$ and $p$:
\begin{align}\label{eq:qpseparable}
	\frac{1}{U'(q) \chi^q_\ve(q)} \frac{d \chi^q_\ve(q)}{dq} - \frac{i}{\hbar} f_\ve(q) =
	\frac{1}{K'(p) \chi^p_\ve(p)} \frac{d \chi^p_\ve(p)}{dp} - \frac{i}{\hbar} g_\ve(p) =
	-\frac{1}{2}\beta_\ve~,
\end{align}
where $-\frac{1}{2}\beta_\ve$ is the separation constant whose significance is discussed below.  Setting the right-hand-side of ~\eqref{eq:qpalmost} is possible only due to the existence of the arbitrary function $\alpha(q,p)$ there.  Because of the freedom to choose $\alpha$, we set it to be
\begin{align}\label{eq:alphaepsilon}
	\alpha(q,p) = \ve - U'(q) K'(p) \left[
		f_\ve(q) - g_\ve(p)
		\right]
\end{align}
to achieve the desired separable form for the equation.  Aside from being real, the functions $f_\ve(q)$ and $g_\ve(p)$ are arbitrary: their form or value will not affect the separability of Eq.~\eqref{eq:qpseparable}.  It is worth emphasizing that such separability would not have been available for equations were $\alpha$ set to zero (as in the Liouville equation or the traditional KvN equation).  Neither would it have been available were $\alpha$ set to $(p \partial_p H - H)$ as considered in Refs.~\cite{Klein2017, Bondar2019}.

The $(q,p)$-separable, time-independent KvN equation~\eqref{eq:qpseparable} has the separable solution
\begin{align}\label{eq:chisep}
	\chi_\ve = \frac{1}{\sqrt{Z^q}} e^{
		-\frac{1}{2}\beta_\ve U(q) + \frac{i}{\hbar} \int f_\ve(q) U'(q) dq
	}
		\cdot \frac{1}{\sqrt{Z^p}} e^{
		-\frac{1}{2}\beta_\ve K(p) + \frac{i}{\hbar} \int g_\ve(p) K'(p) dp
	}
	\cdot e^{-\frac{i}{\hbar} \ve t}~.
\end{align}
Here $Z^q$ and $Z^p$ are normalization constants given by
\begin{align}
	Z^q = \int_{\underline{q}}^{\overline{q}} dq \, e^{-\beta_\ve U(q)},
	\quad
	Z^p = \int_{\underline{p}}^{\overline{p}} dp \, e^{-\beta_\ve K(p)}
\end{align}
where $(\underline{q}, \overline{q})$ and $(\underline{p}, \overline{p})$ are the lower and upper bounds on $q$ and $p$ respectively.  With $\Gamma$ as shorthand for the phase-space region of the problem and $d\Omega$ as the volume element in phase space, $Z(\beta_\ve, \Gamma)$ is defined as\footnote{
	We use $d\Omega$ instead of $dq \, dp$ as a reminder that, due to Liouville's theorem~\cite{Sudarshan2016}, the relation is valid in any choice of canonical coordinates, not only $q$ and $p$.  This is relevant for the discussion in sections~\ref{sec:superposition} and~\ref{sec:orthohermit}.
}
\begin{align}
	Z(\beta_\ve, \Gamma) = Z^q \, Z^p = \int_\Gamma d\Omega \, e^{-\beta_\ve H}~.
\end{align}
This, of course, is the partition function for the $\chi_\ve$ state.

We see that the $(q,p,t)$-separable solution~\eqref{eq:chisep} corresponds to the canonical ensemble probability distribution
\begin{align}\label{eq:canens}
	|\chi_\ve|^2 = \frac{1}{Z(\beta_\ve, \Gamma)} e^{-\beta_\ve H}~,
\end{align}
and that $\beta_\ve$ is related to the average energy in the $\chi_\ve$ state via
\begin{align}
	\langle H \rangle_{\chi_\ve}
	= -\frac{\partial}{\partial \beta_\ve} \log Z(\beta_\ve, \Gamma)
\end{align}
as is well known from statistical mechanics.  In light of these results, the separation constant $\beta_\ve$ can be identified as the reciprocal of ``temperature'' (times the Boltzmann constant).

Temperature in this context is an information property, it arises purely from the probability distribution, even for a single particle.  Recall that, in Jaynes's information-theoretic statistical mechanics~\cite{jaynes_information_1957}, the canonical ensemble emerges as the probability distribution of maximum entropy (ignorance) apart from knowing the average energy.  The distribution is parameterized by $\beta_\ve$ which, in that derivation, is a Lagrange multiplier.

Mathematically, it is not surprising that $(q, p)$-separability produces the canonical ensemble distribution.  We see that the gauged KvN equation is first order in $q$ and $p$, thus separability in $q$ and $p$ produces an exponential function~\eqref{eq:chisep}, giving rise to the canonical ensemble~\eqref{eq:canens}.

Of course, \textit{any} equilibrium solution would be valid---canonical ensemble or otherwise---for \textit{any} time-independent Hamiltonian, separable or not.  As mentioned before, these solutions would depend on physical and statistical arguments.  The above simple derivation of the canonical ensemble distribution, however, is a direct consequence of the search for separable solutions of the gauged KvN equation.

The separability of $q$ and $p$ required choosing $\alpha$ in an $\ve$-dependent way, i.e., every such solution requires a different gauge.  We will discuss how to superpose such solutions in Sec.~\ref{sec:superposition}.  Let us first derive another type of solution that does not require an $\ve$-dependent choice of $\alpha$.  Given a Hamiltonian, one can define a variable $\tau$ to be its Poisson canonical conjugate ($\{\tau, H\} = 1$).\footnote{
	The functional form of $\tau(q,p,t)$ can be found by solving for $\tau$ in the equation $\{\tau, H\}=1$.  Here are some simple examples of $(\tau, H)$ pairs
	\begin{align*}
		\begin{array}{lll}
			\text{Free particle} & H = \frac{1}{2m}p^2~, & \tau = m \frac{q}{p}~, \\
			\text{Linear potential} & H = \frac{1}{2m}p^2 - F q~, & \tau = \frac{1}{F} p~, \\
			\text{Harmonic oscillator} & H = \frac{1}{2m}p^2 + \frac{1}{2}m\omega^2 q^2~, & \tau = \frac{1}{\omega} \tan^{-1}{\left(m \omega \frac{q}{p}\right)}~.
		\end{array}
	\end{align*}
	An arbitrary constant can be added to $\tau$.
}
The phase-space variable $\tau$ is sometimes called the dynamical time~\cite{bhamathi_time_2003}.  In the $(\tau, H)$ system of canonical coordinates, we have $\{f(\tau, H), H\} = \partial_\tau f(\tau,H)$.  If $H$ is time-independent, then so is $\tau$.  The time-independent KvN equation~\eqref{eq:timeind} can be written as
\begin{align}
	-i\hbar \, \partial_\tau \chi_\ve + \alpha \chi_\ve = \ve \chi_\ve~.
\end{align}
This equation can be integrated to give
\begin{align}\label{eq:tauHsol}
	\chi_\ve(\tau, H, t) = f_\ve(H) e^{\frac{i}{\hbar}
		\left[ -\int \alpha(\tau,H) d\tau + \ve (\tau - t) \right]~,
	}
\end{align}
for some normalized function $f_\ve(H)$.  Choosing $\alpha = 0$ in~\eqref{eq:tauHsol} simplifies it to
\begin{align}\label{eq:tauHa0}
	\chi_\ve(\tau, H, t) = f_\ve(H) e^{\frac{i}{\hbar} \ve (\tau - t)}~.
\end{align}
This solution will be convenient for discussing superpositions of $(q,p)$-separable solutions in Sec.~\ref{sec:superposition}, and orthonormal stationary states in Sec.~\ref{sec:orthohermit}.

\section{Cross-gauge superposition}\label{sec:superposition}

In obtaining the $(q,p)$-separable solutions~\eqref{eq:chisep} discussed in the previous section, a crucial choice of $\alpha$ was made in Eq.~\eqref{eq:alphaepsilon}.  That choice is different for different values of $\ve$.  If we wish to superpose these separable states, and to evolve the superposed state in time, we need to find a suitably gauged KvN equation for it.  This guarantees that the absolute square of the superposition is a solution to the Liouville equation.

Let us write a set of gauged KvN equations~\eqref{eq:gaugedKvN}
\begin{align}
	D \chi_\ve = \frac{1}{i\hbar} \alpha_\ve \chi_\ve
\end{align}
where $D := \partial_t - \{H, \cdot\}$ and $\chi_\ve$ is the solution with $\alpha_\ve$.  For simplicity, $\ve$ is discrete, extension to a continuous index is straightforward.  It is easy to see that a naive superposition of $\chi_\ve$ is not a solution to any gauged KvN equation
\begin{align}
	D \sum_\ve c_\ve \chi_\ve = \frac{1}{i\hbar} \sum_\ve \alpha_\ve c_\ve \chi_\ve~,
\end{align}
unless all $\alpha_\ve$ were equal.  This precludes the separation of variables technique and defeats our purpose.

Instead of directly superposing the states $\chi_\ve$, we gauge-transform the pairs $(\chi_\ve, \alpha_\ve)$ to $(\chi_\ve^{(s)}, \alpha^{(s)})$ according to~\eqref{eq:transformation}
\begin{align}\label{eq:supertransformation}
	\begin{split}
		\chi_\ve \to \chi_\ve^{(s)} = \chi_\ve e^{\frac{i}{\hbar} \varphi_\ve}~, \quad
		\alpha_\ve \to \alpha^{(s)} = \alpha_\ve - D \varphi_\ve~.
	\end{split}
\end{align}
Here $\chi_\ve^{(s)}$ are the ``superposable'' states and $\alpha^{(s)}$ is the superposition gauge.  Note that all $\alpha_\ve$ are transformed into the same $\alpha^{(s)}$ in order to produce a single KvN equation for the superposition
\begin{align}
	i\hbar \, \partial_t \left(
		\sum_\ve c_\ve \chi_\ve^{(s)}
	\right) = \wt{H}^{(s)} \left(
		\sum_\ve c_\ve \chi_\ve^{(s)}
	\right)~,
\end{align}
where the superposition tilde-Hamiltonian $\wt{H}^{(s)}$ is $i\hbar \{H, \cdot\} + \alpha^{(s)}$ and $c_\ve$ are constant coefficients.  Equations~\eqref{eq:supertransformation} are the conditions for constructing cross-gauge superposable states.

As a specific example of constructing superposable state out of $(q,p)$-separable ones, we will consider simple choices for the gauges $\alpha_\ve$ and $\alpha^{(s)}$.  The role of $\alpha_\ve$ in Eq.~\eqref{eq:qpalmost} is to help find $(q,p)$-separable solutions, and for that, it suffices to set $\alpha_\ve = \ve$ to give
\begin{align}
	\chi_\ve = \frac{1}{\sqrt{Z(\beta_\ve, \Gamma)}} e^{
		-\frac{1}{2}\beta_\ve H - \frac{i}{\hbar} \ve t
	}~.
\end{align}
The conditions~\eqref{eq:supertransformation} for this system now gives $\alpha^{(s)} = \ve - D \varphi_\ve$.  Once more, for simplicity, we choose to set $\alpha^{(s)} = 0$, like we did to obtain Eq.~\eqref{eq:tauHa0}.  This leaves us with the equation for $\varphi_\ve$ becoming
\begin{align}
	\partial_t \varphi_\ve + \{\varphi_\ve, H\} = \ve~.
\end{align}
As with the choices for $\alpha_\ve$ and $\alpha^{(s)}$, we look for a simple solution for $\varphi_\ve$.  Since $\{\tau, H\}=1$, one choice of $\varphi_\ve$ that solves conditions~\eqref{eq:supertransformation} for this system is $\varphi_\ve = \ve \tau$, so that we arrive at the superposable states
\begin{align}\label{eq:simplechi}
	\chi_\ve^{(s)} = \frac{1}{\sqrt{Z(\beta_\ve, \Gamma)}} e^{-\frac{1}{2}\beta_\ve H} e^{
		\frac{i}{\hbar} \ve (\tau - t)
	}~.
\end{align}
Compare this to~\eqref{eq:tauHa0}.  With the choice $\alpha^{(s)}=0$, the superposition tilde-Hamiltonian $\wt{H}^{(s)} = i\hbar\{H,\cdot\}$ is the ordinary Liouvillian operator, and the states~\eqref{eq:simplechi} constitute its eigenfunctions with eigenvalues $\ve$.

Note that the simplicity of working with states~\eqref{eq:simplechi} is countered by the difficulty of finding the functional form of $\tau(q,p,t)$.  If one needs that form, one has to solve the differential equation $\partial_q \tau \partial_p H - \partial_q H \partial_p \tau = 1$, which might be difficult, depending on the Hamiltonian.  That said, useful results are found by formally using the canonical pair $(\tau, H)$ regardless of their functional dependence on $(q,p,t)$, as will be seen in the rest of this paper.

\section{Hermiticity of the Liouvillian}\label{sec:orthohermit}

One of the attractive features of a Hilbert space framework is the ability to expand any function in the space in terms of eigenfunctions of Hermitian operators.
In this section,
we discuss the Hermiticity of the Liouvillian operator,
or its generalization $\wt{H}$,
and show one way of expanding an arbitrary probability amplitude in terms of its eigenfunctions.
In certain situations,
the set of eigenfunctions of $\wt{H}$ is found to be discrete.

First,
we recognize that the Hilbert space for a classical system of one degree of freedom,
is composed of \textit{two} Hilbert spaces.
For instance,
if the system is described using $q$ and its Poisson-conjugate $p$,
then there is one Hilbert space whose elements are square integrable functions of $q$,
and another for $p$.
This is a reflection of the commutativity of phase-space variables in classical mechanics.

In this section,
we are concerned with the Liouvillian operator
$\wt{H} = i\hbar \{H, \cdot\} + \alpha$
for a system of one degree of freedom.
It is then convenient to work in the $(\tau, H)$ representation.
Since $\{\tau, H\} = 1$,
we have $\wt{H} = -i\hbar \partial_\tau + \alpha$,
where $\alpha(\tau, H)$ is an arbitrary function as discussed in Sec.~\ref{sec:prob-amp}.
In this representation,
a general state amplitude may be written as a linear combination of product states
\begin{align}
	\chi(\tau, H) = \sum_n c_n \, \chi^{(1)}_n(\tau) \chi^{(2)}_n(H)~.
\end{align}
We will now choose the arbitrary gauge $\alpha$ to be a function of $\tau$ only,
so that the operator $\wt{H}$ acts only on the first Hilbert space,
whose functions are marked with the superscript $(1)$.
Now we have operators such as $\tau$ and $\wt{H}=-i\hbar\partial_\tau + \alpha(\tau)$
act on square-integrable functions $\chi^{(1)}(\tau)$,
while $H$ acts on $\chi^{(2)}(H)$.

We may view the eigenvalue equation for $\wt{H}$ as being on the first Hilbert space
\begin{align}
	\wt{H} \phi^{(1)}_\ve(\tau) = \ve \phi^{(1)}_\ve(\tau)
	\quad \Rightarrow \quad
	\phi^{(1)}_\ve(\tau) = A_\ve \, e^{\frac{i}{\hbar} \left[
		\ve \tau - \sigma(\tau)
	\right]}~,
\end{align}
where $\sigma(\tau) = \int \alpha(\tau) \, d\tau$ is real,
$A_\ve$ are normalization constants,
and $\phi^{(1)}_\ve$ are the eigenfunctions of $\wt{H}$ on the first Hilbert space with eigenvalues $\ve$.
We see that the classical $\tau$ and
$\wt{H} = -i\hbar \, \partial_\tau + \alpha$
bear resemblance to the quantum position and momentum.
On this space,
an inner product between two arbitrary functions is given by
\begin{align}
	\langle \chi^{(1)}_a | \chi^{(1)}_b \rangle =
	\int_{\underline{\tau}}^{\overline{\tau}} d\tau \,
	\chi^{*(1)}_a(\tau) \, \chi^{(1)}_b(\tau)~.
\end{align}
The Liouvillian $\wt{H}$ is Hermitian if the boundary conditions are such that
\begin{align}\label{eq:hermiticity-condition}
	\langle \chi^{(1)} | \wt{H} \chi^{(1)} \rangle -
	\langle \wt{H} \chi^{(1)} | \chi^{(1)} \rangle =
	-i\hbar \int_{\underline{\tau}}^{\overline{\tau}} d\tau \,
	\partial_\tau \left| \chi^{(1)}(\tau) \right|^2 = 0
\end{align}
for all functions $\chi^{(1)}$ in the space.
That is,
$\wt{H}$ is Hermitian if
$\chi^{(1)}(\overline{\tau}) = \chi^{(1)}(\underline{\tau}) e^{i\theta}$
for some real constant $\theta$.

In the case that $(\underline{\tau}, \overline{\tau}) = (-\infty, \infty)$,
we must have $\chi^{(1)}(\overline{\tau}) = \chi^{(1)}(\underline{\tau}) = 0$
and $\wt{H}$ is Hermitian
(even though its ``eigenfunctions'' are not in the Hilbert space~\cite{byron_mathematics_1992, ballentine_quantum_2014}),
and we have $\langle \phi^{(1)}_{\ve'} | \phi^{(1)}_\ve \rangle = \delta(\ve - \ve')$.
Since we can expand any function in a Fourier integral as
\begin{align}
	f(\tau) = \int_{-\infty}^{\infty} d\ve \, g(\ve) \,
	\frac{e^{\frac{i}{\hbar} \ve \tau}}{\sqrt{2\pi\hbar}}~,
	\quad
	g(\ve) = \int_{-\infty}^{\infty} d\tau \, f(\tau) \,
	\frac{e^{-\frac{i}{\hbar} \ve \tau}}{\sqrt{2\pi\hbar}}~,
\end{align}
then by writing
$\chi^{(1)}(\tau) = f(\tau) \, e^{-\frac{i}{\hbar} \sigma(\tau)}$
we see that that any function $\chi^{(1)}(\tau)$ can be expanded in terms of
$\phi^{(1)}_\ve(\tau) = \frac{1}{\sqrt{2\pi\hbar}} e^{\frac{i}{\hbar} \left[
	\ve \tau - \sigma(\tau)
\right]}$
as
\begin{align}\label{eq:complete-continuous}
	\chi^{(1)}(\tau) = \int_{-\infty}^{\infty} d\ve \,
	\langle \phi^{(1)}_\ve | \chi^{(1)} \rangle \, \phi^{(1)}_\ve(\tau)~.
\end{align}
In this sense,
the eigenfunctions
$\phi^{(1)}_\ve(\tau)$
form a complete set.

In the case that both $\underline{\tau}$ and $\overline{\tau}$ are finite,
$\wt{H}$ would be Hermitian if
$\chi^{(1)}(\overline{\tau}) = \chi^{(1)}(\underline{\tau}) \, e^{i\theta}$
for real $\theta$,
as discussed below equation~\eqref{eq:hermiticity-condition}.
Such is the case for the Harmonic oscillator,
as will be shown in Sec.~\ref{sec:illust}.
The eigenfunctions
$\phi^{(1)}_\ve(\tau) = \frac{1}{\sqrt{2\pi\hbar}} e^{\frac{i}{\hbar} \left[
	\ve \tau - \sigma(\tau)
\right]}$
belong to the Hilbert space if
$\phi^{(1)}_\ve(\overline{\tau}) = \phi^{(1)}_\ve(\underline{\tau}) \, e^{i\theta}$.
This implies that the spectrum $\ve$ is discrete:
\begin{align}\label{eq:spectrum-discrete}
	\ve_n = \ve_0
	+ \frac{2\pi\hbar}{\overline{\tau} - \underline{\tau}} n~,
	\quad
	\ve_0 = \frac{1}{\overline{\tau} - \underline{\tau}} \left[
		\sigma(\overline{\tau}) - \sigma(\underline{\tau})
		+ \theta \hbar
	\right]~,
	\quad
	n \in \mathbb{Z}~.
\end{align}
The eigenfunctions
\begin{align}
	\phi^{(1)}_n(\tau)
	= \frac{1}{\sqrt{\overline{\tau} - \underline{\tau}}}
		e^{\frac{i}{\hbar} \left[
				\ve_n \tau - \sigma(\tau)
		\right]}
\end{align}
are orthonormal,
$\langle \phi^{(1)}_{n'} | \phi^{(1)}_{n} \rangle = \delta_{n, n'}$.
Similarly to Eq.~\eqref{eq:complete-continuous},
since any function $f(\tau)$ in the space can be expanded in a Fourier series as a linear combination of
$\frac{1}{\sqrt{\overline{\tau} - \underline{\tau}}}
e^{\frac{i}{\hbar} \ve_n \tau }$,
we can prove that
\begin{align}\label{eq:complete-discrete}
	\chi^{(1)}(\tau) = \frac{1}{\sqrt{\overline{\tau} - \underline{\tau}}}
	\sum_{n=-\infty}^{\infty} \langle \phi^{(1)}_n | \chi^{(1)} \rangle \,
	\phi^{(1)}_n(\tau)~,
\end{align}
and so $\phi^{(1)}_n(\tau)$ form a complete set of orthonormal basis functions for the space.


For the second Hilbert space,
made of square-integrable functions of $H$
(marked with the superscript $(2)$),
the inner product is given by
\begin{align}
	\langle \chi^{(2)}_a | \chi^{(2)}_b \rangle =
	\int_{\underline{H}}^{\overline{H}} dH \, \chi^{*(2)}_a(H) \, \chi^{(2)}_b(H)~.
\end{align}
Here we have the ``eigenvalue'' equation for $H$,
whose ``eigenfunctions'' are Dirac delta functions
\begin{align}
	H \phi^{(2)}_E(H) = E \phi^{(2)}_E(H)
	\quad \Rightarrow \quad
	\phi^{(2)}_E(H) = \delta(H - E)~,
\end{align}
as expected from a multiplicative operator,
with $\underline{H} \leq E \leq \overline{H}$.
(Once again,
the functions $\phi^{(2)}_E(H)$ do not belong to the Hilbert space,
but they are useful nonetheless~\cite{byron_mathematics_1992, ballentine_quantum_2014}.)
Any function $\chi^{(2)}(H)$ on the space can be expanded in terms of $\phi^{(2)}_E(H)$ as
\begin{align}
	\chi^{(2)}(H) = \int_{-\infty}^{\infty} dE \,
	\langle \phi^{(2)}_E | \chi^{(2)} \rangle \, \phi^{(2)}_E(H)~.
\end{align}
This is simply the trivial statement that
\begin{align}
	\chi^{(2)}(H) = \int_{-\infty}^{\infty} dE \,
	\int_{\underline{H}}^{\overline{H}} dH \,
	\delta(H - E) \, \chi^{(2)}(H) \, \delta(H - E)~.
\end{align}

The full Hilbert space is composed of square-integrable functions $\chi(\tau, H)$,
with an inner product
\begin{align}
	\langle \chi_a | \chi_b \rangle =
	\int_{\underline{\tau}}^{\overline{\tau}} d\tau \,
	\int_{\underline{H}}^{\overline{H}} dH \,
	\chi_a^*(\tau,H) \, \chi_b(\tau,H)~.
\end{align}
With the complete set of eigenfunctions $\phi^{(1)}_\ve$ of the Hermitian operator $\wt{H}$ on the first Hilbert space,
and the complete set of eigenfunctions $\phi^{(2)}_E$ of $H$ on the second Hilbert space,
we can expand any function on the total Hilbert space.
From the examples above,
if $(\underline{\tau}, \overline{\tau}) = (-\infty, \infty)$,
and the spectrum of $\wt{H}$ is continuous,
we can write
\begin{align}
	\chi(\tau, H)
	&= \int_{-\infty}^{\infty} d\ve \,
	\int_{-\infty}^{\infty} dE \,
	\langle \phi^{(1)}_\ve \, \phi^{(2)}_E | \chi \rangle \,
	\phi^{(1)}_\ve(\tau) \, \phi^{(2)}_E(H)\\
	&= \int_{-\infty}^{\infty} d\ve \,
	f_\ve(H) \,
	\frac{ e^{\frac{i}{\hbar}\left[
				\ve \tau - \sigma(\tau)
			\right]
		}
	}{\sqrt{2\pi\hbar}}
\end{align}
where
\begin{align}
	f_\ve(H)
	&= \int_{-\infty}^{\infty} dE \,
	\langle \phi^{(1)}_\ve \, \phi^{(2)}_E | \chi \rangle \,
	\phi^{(2)}_E(H)\\
	&= \int_{-\infty}^{\infty} d\tau \,
	\chi(\tau, H) \,
	\frac{ e^{-\frac{i}{\hbar}\left[
				\ve \tau - \sigma(\tau)
			\right]
		}
	}{\sqrt{2\pi\hbar}}~.
\end{align}
If $(\underline{\tau}, \overline{\tau})$ are finite,
and the spectrum of $\wt{H}$ is discrete,
we similarly have
\begin{align}
	\chi(\tau, H)\label{eq:chi-discrete}
	&= \sum_{n=-\infty}^{\infty} f_n(H) \,
	\frac{ e^{\frac{i}{\hbar}\left[
				\ve_n \tau - \sigma(\tau)
			\right]
		}
	}{\sqrt{\overline{\tau} - \underline{\tau}}}~,\\
	f_n(H) \label{eq:fn-discrete}
	&= \int_{\underline{\tau}}^{\overline{\tau}} d\tau \,
	\chi(\tau, H) \,
	\frac{ e^{-\frac{i}{\hbar}\left[
				\ve_n \tau - \sigma(\tau)
			\right]
		}
	}{\sqrt{\overline{\tau} - \underline{\tau}}}~.
\end{align}

The general probability amplitude formed through superposition such as above,
we find the Liouville probability distribution to be
\begin{align}\label{eq:probdist-continuous}
	\rho(\tau, H) =
	\iint_{-\infty}^{\infty} d\ve \, d\ve' \,
	f^*_{\ve'}(H) \, f_\ve(H) \,
	\frac{ e^{\frac{i}{\hbar} (\ve - \ve') \tau}
	}{2\pi\hbar}~,
\end{align}
for $(\underline{\tau}, \overline{\tau}) = (-\infty, \infty)$,
or
\begin{align}\label{eq:probdist-discrete}
	\rho(\tau, H) =
	\sum_{n, n' = -\infty}^{\infty}
	f^*_{n'}(H) \, f_n(H) \,
	\frac{ e^{\frac{i}{\hbar} (\ve_n - \ve_{n'}) \tau}
	}{\overline{\tau} - \underline{\tau}}~,
\end{align}
for both $\underline{\tau}$ and $\overline{\tau}$ finite.
We see that that the differences in eigenvalues of the tilde-Hamiltonian appear as parameters defining the probability distribution,
and that the gauge-dependent phase
$\sigma(\tau) = \int \alpha(\tau) \, d\tau$
has dropped out,
as expected.

Since the equation of motion is
$i\hbar \, \partial_t \chi = \wt{H} \chi$,
the time dependence of the eigenfunctions $\phi^{(1)}_\ve$
(either discrete or continuous)
is given by
\begin{align}
	\phi^{(1)}_\ve(\tau, t) =
	A_\ve \, e^{\frac{i}{\hbar} \left[\ve (\tau - t) - \sigma(\tau)\right]}~.
\end{align}
The time-evolution of probability distributions~\eqref{eq:probdist-continuous} and~\eqref{eq:probdist-discrete} is then given by
\begin{align}\label{eq:probdist-evolution}
	\rho(\tau, H, t) = \rho(\tau - t, H)~.
\end{align}
That is,
the time-evolved probability distribution is the initial distribution with its dynamical-time $\tau$ shifted by the amount of time $t$.

\section{The simple harmonic oscillator}\label{sec:illust}

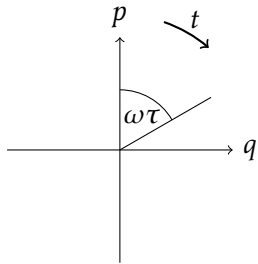
\begin{figure}[h]
	\centering
	\begin{tikzpicture}
		\draw[->] (-1.5, 0) -- (1.5, 0) node[right] {$q$};
		\draw[->] (0, -1.5) -- (0, 1.5) node[above] {$\frac{1}{m\omega}p$};
		\draw[thick, ->] (0.586, 1.702) arc (71:49:1.8);
		\draw (1, 1.732) node {$t$};
		\draw (0, 0) -- (1.212, 0.7);
		\draw (0, 0.8) arc (90:30:0.8);
		\draw (0.3, 0.45) node {$\omega\tau$};
	\end{tikzpicture}
	\caption{
		For the simple harmonic oscillator,
		a given $\tau$ defines a ray at an angle $\omega\tau$ clockwise from the $p$-axis in the $qp$-plane.
		Time evolution shifts $\tau$ by an amount $t$.
	}
	\label{fig:tauH}
\end{figure}

Consider the 1D simple harmonic oscillator defined by the Hamiltonian
$H(q,p) = \frac{1}{2m} p^2 + \frac{1}{2} m \omega^2 q^2$.
%
%
To find $\tau(q,p)$,
we must solve for it in the differential equation
$\{\tau, H\} = 1$.
For the above Hamiltonian, the solution is
$\frac{1}{\omega} \tan^{-1}{\left(m\omega \frac{q}{p}\right)} + c$
for some arbitrary constant $c$.
The $\tan$ function is periodic,
and so $\tan^{-1}$ is multivalued,
unless defined carefully.
One can naively set
$\tau = \frac{1}{\omega}\tan^{-1} \left(m\omega \frac{q}{p}\right)$
with the limits
$(-\frac{\pi}{2\omega}, \frac{\pi}{2\omega})$
but then integrals over $dq\,dp$ are not equal to those over $d\tau \, dH$,
as demanded by Liouville's theorem.
Instead, we set $\tau$ as
\begin{align}\label{eq:tauSHO}
	\tau = \left\{
	\begin{array}{lllc}
		\frac{1}{\omega} \tan^{-1}{\left(m\omega \frac{q}{p}\right)} - \frac{\pi}{\omega}~,
		& q < 0 \text{ and } p < 0~, &\text{or}
		&\frac{-\pi}{\omega} < \tau < \frac{-\pi}{2\omega}~, \\
		\frac{1}{\omega} \tan^{-1}{\left(m\omega \frac{q}{p}\right)}~,
		& p \geq 0~, &\text{or}
		&\frac{-\pi}{2\omega} \leq \tau \leq \frac{\pi}{2\omega}~, \\
		\frac{1}{\omega} \tan^{-1}{\left(m\omega \frac{q}{p}\right)} + \frac{\pi}{\omega}~,
		& q \geq 0 \text{ and } p < 0~, &\text{or}
		&\frac{\pi}{2\omega} < \tau \leq \frac{\pi}{\omega}~.
	\end{array}
	\right.
\end{align}
We see that the limits on $\tau$ are
$(\underline{\tau}, \overline{\tau}) = (-\frac{\pi}{\omega}, \frac{\pi}{\omega})$.
This guaranties that integrals in the $(\tau, H)$ coordinates are equal to those in the $(q,p)$ ones.
The inverse transformation is given by
\begin{align}\label{eq:inverseqp}
	q = \frac{1}{m\omega} \sqrt{2mH} \, \sin{\omega\tau}~,
	\quad
	p = \sqrt{2mH} \, \cos{\omega\tau}~.
\end{align}
On the $qp$-plane,
$\omega\tau$ is the angle,
going clockwise,
between the $p$-axis and a line connecting the origin and a point $(q,p)$,
see Fig.~\ref{fig:tauH}.
Time evolution shifts the dynamical time $\tau$ by amount $t$,
which increases the angle $\omega\tau$.

Since probability distributions are single-valued,
we must have
$\rho(\overline{\tau}, H) = \rho(\underline{\tau}, H)$.
Therefore,
the Hilbert space is composed of probability amplitudes obeying
$\chi(\overline{\tau}, H) =
\chi(\underline{\tau}, H) \, e^{i \theta}$
for some real constant $\theta$,
and $\wt{H}$ is a Hermitian operator,
as discussed in Sec.~\ref{sec:orthohermit}.
The eigenvalues of $\wt{H}$ are discrete,
and are given by Eq.~\eqref{eq:spectrum-discrete} by
\begin{align}
	\ve_n = \ve_0 + n \hbar \omega~,
	\quad
	n = 0, \pm 1, \pm 2, \cdots~.
\end{align}

We see that the classical simple harmonic oscillator has some similarities with its quantum counterpart.
In both theories,
the generator of time-evolution has a discrete spectrum,
and the difference between two successive values of the spectrum is $\hbar\omega$.
On the other hand,
the two theories differ in that
the generator of time evolution in quantum mechanics is itself the energy of the system.
In the case of the harmonic oscillator,
that energy is always positive,
implying that the spectrum is bounded from below
($n = 0, 1, 2, \cdots$).
Such is not the case in the classical system where the energy is continuous,
and is not identified with the generator of time evolution $\wt{H}$.
Further,
the appearance of $\hbar$ in the classical calculation above,
it should be emphasized,
was due to an arbitrary choice made in Eq.~\eqref{eq:gaugedKvN},
and does not appear to have fundamental significance.

Let us now consider an example of a simple probability amplitude.
The treatment here is applicable to any system with periodic boundary conditions for the probability distribution over $\tau$,
including the harmonic oscillator.
Let the initial probability amplitude take the form
\begin{align}\label{eq:chi-initial}
	\chi(\tau,H) =
	\begin{cases}
		\frac{1}{\sqrt{\Delta \tau}} \, \chi^{(2)}(H)~,
		& |\tau - \tau_i| \leq \frac{1}{2} \Delta \tau~,\\
		0~, & \text{otherwise}~.
	\end{cases}
\end{align}
where $\chi^{(2)}(H)$ is some square-integrable function over $H$.
%
%
From Eq.~\eqref{eq:chi-discrete},
the initial probability amplitude is
expanded in terms of eigenfunctions of $\wt{H}$,
is given by
$ \sum_n f_n(H) \,
\frac{ e^{\frac{i}{\hbar}\left[
			\ve_n \tau - \sigma(\tau)
		\right]
	}
}{\sqrt{\overline{\tau} - \underline{\tau}}} $.
The functions $f_n(H)$ are given by Eq.~\eqref{eq:fn-discrete}
and,
in our case,
become
\begin{align}
	f_n(H)
	= \frac{1}{\sqrt{\Delta\tau}} \chi^{(2)}(H)
	\int_{\tau_i - \frac{1}{2}\Delta\tau}^{\tau_i + \frac{1}{2}\Delta\tau} d\tau \,
	\frac{ e^{-\frac{i}{\hbar} \left[ \ve_n \tau - \sigma(\tau) \right]}
	}{\sqrt{\overline{\tau} - \underline{\tau}}}~.
\end{align}
Evaluating this integral requires setting the arbitrary function
$\sigma(\tau) = \int \alpha(\tau) \, d\tau$
first.
One simple choice is to set $\alpha(\tau) = \ve_N$,
where $\ve_N$ is an arbitrary eigenvalue of $\wt{H}$ with some integer $N$,
so that $\sigma(\tau) = \ve_N \tau$.
Now the integral simply evaluates to
\begin{align}
	f_n(H) = 
	\begin{cases}
		\sqrt{\frac{\Delta\tau}{\overline{\tau}-\underline{\tau}}} \, \chi^{(2)}(H)~,
		& n = N~,\\
		\sqrt{\frac{\overline{\tau}-\underline{\tau}}{\Delta\tau}} \,
		\frac{ e^{-i\frac{2\pi(n-N)}{\overline{\tau}-\underline{\tau}}\tau_i}
		}{\pi(n-N)} \,
		\sin \left[\frac{\pi(n-N)}{\overline{\tau}-\underline{\tau}}\Delta\tau\right] \,
		\chi^{(2)}(H)~,
		& n \neq N~.
	\end{cases}
\end{align}

We see that if
$\Delta\tau = \overline{\tau}-\underline{\tau}$,
then all of $f_n(H)$ vanish,
except for $f_N(H) = \chi^{(2)}(H)$,
and the state $\chi(\tau, H)$ becomes an eigenstate of $\wt{H}$
with eigenvalue $\ve_N$.
That is,
an eigenstate of $\wt{H}$ is one of maximal uncertainty in the dynamical time $\tau$.
Conversely,
an eigenstate of $\tau$,
given by
$\chi(\tau, H) = \delta(\tau - \tau_i) \chi^{(2)}(H)$
(or informally, by taking $\Delta\tau \to 0$ in Eq.~\eqref{eq:chi-initial}).
This state expands to
$ \sum_n \,
\frac{ e^{i \frac{2\pi n}{\overline{\tau}-\underline{\tau}}(\tau - \tau_i)}
}{\overline{\tau} - \underline{\tau}} \chi^{(2)}(H) $.\footnote{
	This can be proved using the properties
	$\delta(x) = \frac{1}{2\pi} \sum_{n=-\infty}^{\infty} e^{i n x}$
	and $\delta(a x) = \frac{1}{a} \delta(x)$
	of the Dirac delta function,
	by setting
	$x = \frac{2\pi}{\overline{\tau}-\underline{\tau}} (\tau - \tau_i)$
	and $a = \frac{\overline{\tau}-\underline{\tau}}{2\pi}$.
}
It is easy to show that this sum runs over all the eigenstates of $\wt{H}$
(with $\alpha(\tau) = \ve_N$).
Since the coefficients in the sum are independent of $n$,
all of the eigenstates contribute equally,
which represents a state of maximal uncertainty in $\wt{H}$.
This is explored further in the next section.

\section{Liouvillian, uncertainty and collapse}\label{sec:LUC}

The KvN framework, as in Eq.~\eqref{eq:SchKvN}, uses the tilde-Hamiltonian $\wt{H}$ for time evolution.  The tilde-Hamiltonian, however, cannot be constructed only from the familiar variables $q$ and $p$ (or any canonical pair).  Since this is a classical theory, $q$ and $p$ commute with each other.  The operator $\wt{H}$ does not commute with either of them, and so must be constructed from other operators.  See~\cite{Gozzi2004} for a detailed discussion.

Define Hermitian operators $\wt{q}$ and $\wt{p}$ such that\footnote{
	The operators $\wt{q}$ and $\wt{p}$ appear in~\cite{Mauro2002, Gozzi2004, Bondar2012, Wilczek2015} in a slightly different form related to ours through simple rescaling as $ \lambda_q = \wt{p} / \hbar$ and $\lambda_p = - \wt{q} / \hbar$.  The appearance of $\hbar$ in our definition gives $\wt{q}$ and $\wt{p}$ the dimensions of $q$ and $p$, respectively.
}
\begin{align}
	\frac{1}{i\hbar} [q, \wt{p}]
	= \frac{1}{i\hbar} [\wt{q}, p] = 1~,
\end{align}
and $[q,\wt{q}] = [p,\wt{p}] = 0$.  In the $(q, p)$ representation, we can write
\begin{align}
	\wt{q} = i\hbar \, \partial_p~, \quad
	\wt{p} = - i\hbar \, \partial_q~, \quad \text{and} \quad
	\wt{H} = \partial_q H \, \wt{q} + \partial_p H \, \wt{p} + \alpha(q,p,t)~.
\end{align}
The existence of such operators might seem surprising, but they have been lurking in the structure of classical mechanics all along.  Indeed, the Liouvillian operator is nothing but a special case of the tilde-Hamiltonian when $\alpha=0$.

 It is natural to wonder whether the ``tilde-variables'' (like $\wt{q}$, $\wt{p}$, and $\wt{H}$) appearing in this formulation are classical.  In~\cite{Wilczek2015}, Wilczek asserts that these extra variables go beyond classical mechanics.  There, he argues that while $\rho = |\chi|^2$ is classically meaningful, the phase of $\chi$ is not, and so ``[o]nly observables which do not depend on that phase choice will be meaningful, classically.''  According to this criterion, the tilde-Hamiltonian $\wt{H}$ (or the Liouvillian, when $\alpha=0$) is not a classical observable, it is gauge-dependent.  In~\cite{Gozzi2004}, Gozzi and Mauro conclude that ``the KvN theory is a generalization of [classical mechanics].''  They invoke a superselection principle so that ``[phase-space] observables could never detect the relative phases contained in [a] coherent superposition.''

 If the Koopman-von Neumann theory, with its noncommutative variables and probability amplitudes is more than classical, but certainly less than quantum, then what is it?  It may be the case that this theory is nothing more than classical mechanics with uncertainty, or simply, statistical mechanics.  That is, a theory of real physical events \textit{and} our knowledge of such events.  Here, we will attempt to explain the tilde-Hamiltonian $\wt{H}$ in statistical terms.  For a general analysis of tilde-variables, their meaning, algebraic properties, and the ``quantum-like'' behaviour they give rise to, see Ref.~\cite{amin_uncertainty_2026}.
 
Let us discuss one of the seemingly ``beyond classical'' features of the KvN theory.  Since $\tau$ and $H$ obey the canonical Poisson relation $\{\tau, H\} = 1$, then $\tau$ and $\wt{H}$ obey the canonical commutation relation $\frac{1}{i\hbar} [\tau, \wt{H}] = 1$, implying an uncertainty trade-off between the dynamical time $\tau$ and the gauged Liouvillian/tilde-Hamiltonian $\wt{H}$
\begin{align}\label{eq:uncertainty}
	\Delta \tau \Delta \wt{H} \geq \frac{\hbar}{2}~.
\end{align}
In other words, eigenstates of $\wt{H}$ (equilibrium states, as we have seen) are states of complete ignorance of $\tau$.  To see the significance of this, consider a 1D system.  Since $(\tau, H)$ are a Poisson-canonical pair, they suffice to completely specify the state of the system on phase space.  If $\tau$ is unknown, then half the information about that system is missing.  More specifically, since $\tau$ is the dynamical time, it serves as an internal clock for the system~\cite{bhamathi_time_2003}, but in a state of equilibrium (eigenstate of $\wt{H}$) there is no meaningful passage of time as far as our knowledge of the system is concerned.

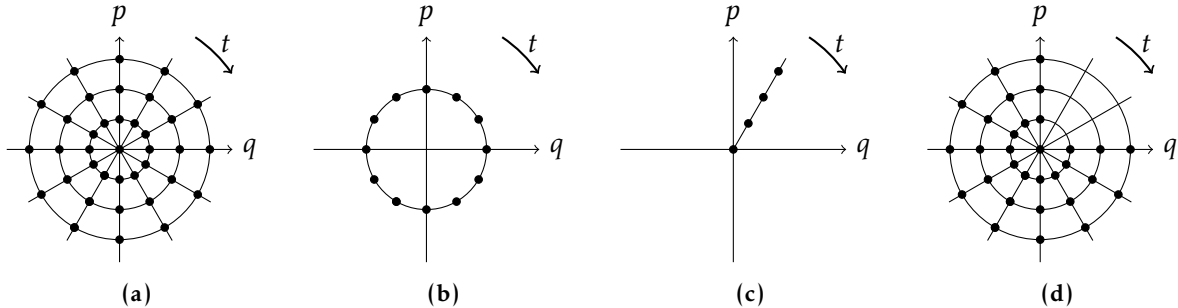
\begin{figure}[h]
	\centering

	\begin{subfigure}{0.24\textwidth}
		\centering
		\begin{tikzpicture}
			\draw[->] (-1.5, 0) -- (1.5, 0) node[right] {$q$};
			\draw[->] (0, -1.5) -- (0, 1.5) node[above] {$\frac{1}{m\omega}p$};
			\draw[thick, ->] (1.007, 1.493) arc (56:34:1.8);
			\draw (1.414, 1.414) node {$t$};

			\draw (-1.212, -0.7) -- (1.212, 0.7);
			\draw (-0.7, -1.212) -- (0.7, 1.212);
			\draw (0.7, -1.212) -- (-0.7, 1.212);
			\draw (1.212, -0.7) -- (-1.212, 0.7);
			\draw (0, 0) circle (0.4);
			\draw (0, 0) circle (0.8);
			\draw (0, 0) circle (1.2);
			
			\filldraw [black] (0, 0) circle (0.05);

			\filldraw [black] (0.4, 0) circle (0.05);
			\filldraw [black] (-0.4, 0) circle (0.05);
			\filldraw [black] (0, 0.4) circle (0.05);
			\filldraw [black] (0, -0.4) circle (0.05);
			\filldraw [black] (0.346, 0.2) circle (0.05);
			\filldraw [black] (0.2, 0.346) circle (0.05);
			\filldraw [black] (-0.2, 0.346) circle (0.05);
			\filldraw [black] (-0.346, 0.2) circle (0.05);
			\filldraw [black] (-0.346, -0.2) circle (0.05);
			\filldraw [black] (-0.2, -0.346) circle (0.05);
			\filldraw [black] (-0.2, -0.346) circle (0.05);
			\filldraw [black] (0.2, -0.346) circle (0.05);
			\filldraw [black] (0.346, -0.2) circle (0.05);

			\filldraw [black] (0.8, 0) circle (0.05);
			\filldraw [black] (-0.8, 0) circle (0.05);
			\filldraw [black] (0, 0.8) circle (0.05);
			\filldraw [black] (0, -0.8) circle (0.05);
			\filldraw [black] (0.694, 0.4) circle (0.05);
			\filldraw [black] (0.4, 0.694) circle (0.05);
			\filldraw [black] (-0.4, 0.694) circle (0.05);
			\filldraw [black] (-0.694, 0.4) circle (0.05);
			\filldraw [black] (-0.694, -0.4) circle (0.05);
			\filldraw [black] (-0.4, -0.694) circle (0.05);
			\filldraw [black] (-0.4, -0.694) circle (0.05);
			\filldraw [black] (0.4, -0.694) circle (0.05);
			\filldraw [black] (0.694, -0.4) circle (0.05);

			\filldraw [black] (1.2, 0) circle (0.05);
			\filldraw [black] (-1.2, 0) circle (0.05);
			\filldraw [black] (0, 1.2) circle (0.05);
			\filldraw [black] (0, -1.2) circle (0.05);
			\filldraw [black] (1.039, 0.6) circle (0.05);
			\filldraw [black] (0.6, 1.039) circle (0.05);
			\filldraw [black] (-0.6, 1.039) circle (0.05);
			\filldraw [black] (-1.039, 0.6) circle (0.05);
			\filldraw [black] (-1.039, -0.6) circle (0.05);
			\filldraw [black] (-0.6, -1.039) circle (0.05);
			\filldraw [black] (-0.6, -1.039) circle (0.05);
			\filldraw [black] (0.6, -1.039) circle (0.05);
			\filldraw [black] (1.039, -0.6) circle (0.05);
		\end{tikzpicture}
		\caption{}
		\label{fig:wH}
	\end{subfigure}
	\begin{subfigure}{0.24\textwidth}
		\centering
		\begin{tikzpicture}
			\draw[->] (-1.5, 0) -- (1.5, 0) node[right] {$q$};
			\draw[->] (0, -1.5) -- (0, 1.5) node[above] {$\frac{1}{m\omega}p$};
			\draw[thick, ->] (1.007, 1.493) arc (56:34:1.8);
			\draw (1.414, 1.414) node {$t$};

			\draw (0, 0) circle (0.8);
			
			\filldraw [black] (0.8, 0) circle (0.05);
			\filldraw [black] (-0.8, 0) circle (0.05);
			\filldraw [black] (0, 0.8) circle (0.05);
			\filldraw [black] (0, -0.8) circle (0.05);
			\filldraw [black] (0.694, 0.4) circle (0.05);
			\filldraw [black] (0.4, 0.694) circle (0.05);
			\filldraw [black] (-0.4, 0.694) circle (0.05);
			\filldraw [black] (-0.694, 0.4) circle (0.05);
			\filldraw [black] (-0.694, -0.4) circle (0.05);
			\filldraw [black] (-0.4, -0.694) circle (0.05);
			\filldraw [black] (-0.4, -0.694) circle (0.05);
			\filldraw [black] (0.4, -0.694) circle (0.05);
			\filldraw [black] (0.694, -0.4) circle (0.05);
		\end{tikzpicture}
		\caption{}
		\label{fig:HwH}
	\end{subfigure}
	\begin{subfigure}{0.24\textwidth}
		\centering
		\begin{tikzpicture}
			\draw[->] (-1.5, 0) -- (1.5, 0) node[right] {$q$};
			\draw[->] (0, -1.5) -- (0, 1.5) node[above] {$\frac{1}{m\omega}p$};
			\draw[thick, ->] (1.007, 1.493) arc (56:34:1.8);
			\draw (1.414, 1.414) node {$t$};

			\draw (0, 0) -- (0.7, 1.212);

			\filldraw [black] (0, 0) circle (0.05);
			\filldraw [black] (0.2, 0.346) circle (0.05);
			\filldraw [black] (0.4, 0.694) circle (0.05);
			\filldraw [black] (0.6, 1.039) circle (0.05);
		\end{tikzpicture}
		\caption{}
		\label{fig:tau}
	\end{subfigure}
	\begin{subfigure}{0.24\textwidth}
		\centering
		\begin{tikzpicture}
			\draw[->] (-1.5, 0) -- (1.5, 0) node[right] {$q$};
			\draw[->] (0, -1.5) -- (0, 1.5) node[above] {$\frac{1}{m\omega}p$};
			\draw[thick, ->] (1.007, 1.493) arc (56:34:1.8);
			\draw (1.414, 1.414) node {$t$};

			\draw (-1.212, -0.7) -- (1.212, 0.7);
			\draw (-0.7, -1.212) -- (0.7, 1.212);
			\draw (0.7, -1.212) -- (-0.7, 1.212);
			\draw (1.212, -0.7) -- (-1.212, 0.7);
			\draw (0, 0) circle (0.4);
			\draw (0, 0) circle (0.8);
			\draw (0, 0) circle (1.2);
			
			\filldraw [black] (0, 0) circle (0.05);

			\filldraw [black] (0.4, 0) circle (0.05);
			\filldraw [black] (-0.4, 0) circle (0.05);
			\filldraw [black] (0, 0.4) circle (0.05);
			\filldraw [black] (0, -0.4) circle (0.05);
			\filldraw [black] (-0.2, 0.346) circle (0.05);
			\filldraw [black] (-0.346, 0.2) circle (0.05);
			\filldraw [black] (-0.346, -0.2) circle (0.05);
			\filldraw [black] (-0.2, -0.346) circle (0.05);
			\filldraw [black] (-0.2, -0.346) circle (0.05);
			\filldraw [black] (0.2, -0.346) circle (0.05);
			\filldraw [black] (0.346, -0.2) circle (0.05);

			\filldraw [black] (0.8, 0) circle (0.05);
			\filldraw [black] (-0.8, 0) circle (0.05);
			\filldraw [black] (0, 0.8) circle (0.05);
			\filldraw [black] (0, -0.8) circle (0.05);
			\filldraw [black] (-0.4, 0.694) circle (0.05);
			\filldraw [black] (-0.694, 0.4) circle (0.05);
			\filldraw [black] (-0.694, -0.4) circle (0.05);
			\filldraw [black] (-0.4, -0.694) circle (0.05);
			\filldraw [black] (-0.4, -0.694) circle (0.05);
			\filldraw [black] (0.4, -0.694) circle (0.05);
			\filldraw [black] (0.694, -0.4) circle (0.05);

			\filldraw [black] (1.2, 0) circle (0.05);
			\filldraw [black] (-1.2, 0) circle (0.05);
			\filldraw [black] (0, 1.2) circle (0.05);
			\filldraw [black] (0, -1.2) circle (0.05);
			\filldraw [black] (-0.6, 1.039) circle (0.05);
			\filldraw [black] (-1.039, 0.6) circle (0.05);
			\filldraw [black] (-1.039, -0.6) circle (0.05);
			\filldraw [black] (-0.6, -1.039) circle (0.05);
			\filldraw [black] (-0.6, -1.039) circle (0.05);
			\filldraw [black] (0.6, -1.039) circle (0.05);
			\filldraw [black] (1.039, -0.6) circle (0.05);
		\end{tikzpicture}
		\caption{}
		\label{fig:taupi}
	\end{subfigure}

	\caption{
		A schematic depiction of harmonic oscillator ensembles.  The solid dots are members of the ensemble.  Circles represent constant $H$ and rays constant $\tau$.  Time-evolution rotates the ensemble clockwise.  \textbf{(a)} An equilibrium distribution remains invariant under time evolution.  \textbf{(b)} An equilibrium distribution with definite energy remains invariant under time evolution.  \textbf{(c)} A state of definite $\tau$ is not invariant under time evolution; it cannot be an equilibrium state.  \textbf{(d)} A state of large, but not infinite, uncertainty in $\tau$, still cannot be an equilibrium state.
	}
	\label{fig:uncertainty}
\end{figure}

Take the 1D harmonic oscillator system as a concrete example of a classical uncertainty relation.\footnote{
	For the harmonic oscillator,
	probability distributions have periodic boundary conditions over $\tau$,
	therefore,
	similar to the situation of angular position and angular momentum for a particle on a circle in quantum mechanics,
	the uncertainty relation~\eqref{eq:uncertainty} is not valid.
	However,
	some nontrivial relation between the uncertainty in $\tau$ and that in $\wt{H}$ exists,
	as discussed in Sec.~\ref{sec:illust} and in this section.
	A formula for the relation may be found,
	for example,
	using entropic uncertainty relations~\cite{coles_entropic_2017},
	but the form of the relation is not essential to the argument presented here.
}
The Hamiltonian is given by $H(q,p) = \frac{1}{2m}p^2 + \frac{1}{2}m\omega^2q^2$, and the dynamical time by $\tau(q,p) = \frac{1}{\omega} \tan^{-1}{\left( m\omega \frac{q}{p} \right)} + c$.  An eigenstate of $\wt{H}$ gives an equilibrium state that does not evolve in time, see Fig.~\ref{fig:wH}.  More precisely, our knowledge of the system, does not.  Since $[H, \wt{H}] = 0$, we can have a simultaneous eigenstate of both $\wt{H}$ and $H$, an equilibrium state of definite energy which is depicted in Fig.~\ref{fig:HwH}~.  On the other hand, since $\tau$ and $\wt{H}$ do not commute, there cannot exist an equilibrium state that has definite $\tau$.  This can be seen since an ensemble with definite $\tau$ at one instant is one where the ratio $q/p$ is definite.  But this constraint defines a ray as shown in Fig~\ref{fig:tau}, such a ray then rotates with time, i.e., it does not represent a state of equilibrium.  Even little information about $\tau$ is enough to forbid equilibrium (Fig.~\ref{fig:taupi}).  Knowledge of $\tau$ and ``knowledge'' of $\wt{H}$ are, quite literally, incompatible.

In hindsight, it is not surprising that such statistical classical uncertainty relations exist.  Since, by assumption, a classical system really occupies a definite point on phase space, any uncertainty is attributed to our knowledge.  However, as we have seen from the example above, it is \textit{not} the case that, given a pair of noncommuting variables, we can only determine one and are merely unable to determine the other with precision.  Rather, if one is sharply defined, then the other is \textit{unsharp}.  This indeed sounds very reminiscent of standard explanations of the quantum uncertainty relations.  We see here that such uncertainty trade-offs are inherent in the statistical description, regardless of the fundamental classical reality of the system described.  It would be interesting to see whether these uncertainty relations can mimic the results of Spekkens's toy theory, where only half the information is available in a classical statistical setup.  The toy theory reproduces some phenomena that are often described as characteristically quantum~\cite{spekkens_defense_2007, bartlett_reconstruction_2012}.

The discussion above frames the uncertainty relation $\Delta \tau \Delta \wt{H} \geq \frac{\hbar}{2}$ entirely in terms of classical statistical mechanics.  It is possible that other uncertainty relations arising from the KvN formalism can be explained similarly.  However, since phase-space variables, by classical assumption, suffice to describe the real state of the system, it is reasonable to believe that the tilde-variables are not on the same footing with phase-space ones.

If we believe that the system is in a state of equilibrium, and if we perform a measurement that determines a value for the dynamical time, then the state collapses from an eigenstate of $\wt{H}$ to one of $\tau$.  This collapse is not accompanied by the conundrums surrounding its quantum counterpart.  It only describes updating our knowledge of the system, not its real, physical state.

Now consider the opposite case.  Say we measure a phase-space variable (like $\tau(q,p)$), and follow that by ``measuring'' an incompatible tilde-variable (like determining that the state is in equilibrium).  If the implication is that we have a collapse to the tilde-variable eigenstate, then the real, physical state of the system did change and a physical disturbance must have occurred.  Such collapse is more of a \textit{preparation} of the system in a state of equilibrium: an eigenstate of $\wt{H}$.

\section{Conclusion}\label{sec:conclusion}

Attempts to understand and pinpoint the crucial differences between classical and quantum mechanics face a major distraction: traditionally, the two theories speak different languages.  It is often thought that Hilbert space is quantum territory, and that quantum ``weirdness'' arises from Hilbert space, where reside superposition and noncommutativity.  Even though classical mechanics was formulated in Hilbert space shortly after quantum mechanics~\cite{Koopman1931, neumann_zur_1932, neumann_zusatze_1932}, this is sometimes viewed as little more than a curiosity.  The KvN formulation allows classical mechanics to have superposition of states and noncommutative operators, but the lack of clear meaning behind these concepts in a classical setting may be the reason this language is not more widely adopted.

It seems that these somewhat ``quantum-like'' features of the Koopman-von Neumann formulation may be attributable to its naturally statistical description of classical mechanics.  Let us recount the connections to statistical mechanics discussed in this paper.  The tilde-Hamiltonian $\wt{H}$ (a gauged Liouvillian) is the operator responsible for time evolution, and it does not commute with phase-space variables. Eigenstates of a time-independent $\wt{H}$ produce, perhaps unsurprisingly, equilibrium distributions; their superposition produce time-dependent ones.  Using the gauge invariance of the KvN equation, one can solve the latter for $(q,p)$-separable solutions where those canonical variables are not correlated, and correlations can be constructed by superposition.  Equilibrium $(q,p)$-separable eigenstates of the tilde-Hamiltonian turn out to produce the canonical ensemble distribution.  Can other distributions from statistical mechanics be derived from eigenstates of other operators?  We are hopeful that a Hilbert-space approach to statistical mechanics may provide a conceptual framework for producing and analyzing different types of distributions.  At present, however, we have not developed such framework.  For example, an analysis of the grand canonical ensemble would require taking a varying number of particles into account, which is not addressed in the present paper.

Uncertainty relations between operators appearing in this classical theory can be understood as an expression of missing knowledge in a given state.  As an example, the relation $\Delta \tau \Delta \wt{H} \geq \frac{\hbar}{2}$ between the dynamical time $\tau$ and the tilde-Hamiltonian $\wt{H}$ implies that an eigenstate of $\wt{H}$ is one where passage of the dynamical time cannot be defined.  Similarly, since we have $\Delta q \Delta \wt{p} \geq \frac{\hbar}{2}$ and $\Delta \wt{q} \Delta p \geq \frac{\hbar}{2}$, we can expect that some ``knowledge'' of $\wt{q}$ implies an uncertainty in $p$, or some width in the distribution of momenta, and similarly for $\wt{p}$ and $q$.  The KvN formulation provides a Hilbert-space framework where our ``state of knowledge'', not just the ``state of reality'' of the system is taken into account, a potentially interesting prospect.\footnote{
	 These are sometimes called ``epistemic'' and ``ontic'' states, respectively
	 (see Ref.~\cite{spekkens_defense_2007}, for example).
}

The existence of ``extra'' noncommutative variables in the KvN theory, and of phases in its probability amplitudes, has been a source of confusion.  Perhaps, to address the question of whether the KvN theory is classical or a generalization that contains nonclassical objects, a modern analysis of the classicality of KvN theory, in terms of noncontextuality and Bell locality, is in order.  It has been argued in~\cite{spekkens_defense_2007, bartlett_reconstruction_2012, spekkens_quasi-quantization_2016, catani_why_2023} that many aspects of phenomena considered essentially quantum can be reproduced in a class of statistical classical theories.  These include ``noncommutativity, interference, collapse, no-cloning, teleportation, remote steering, and entanglement''~\cite{catani_what_2022}.  It would be interesting to investigate these phenomena in KvN classical mechanics, given its builtin mechanisms for uncertainties and superposition.

\section*{Acknowledgements}
We thank Carlo Maria Scandolo for helpful discussion on classical theories with an epistemic restriction.  This research was supported in part by a Discovery Grant (M.A.W.) from the Natural Sciences and Engineering Research Council of Canada (funding reference No. RGPIN-2022-04225).

\AtNextBibliography{\footnotesize}
\printbibliography

\end{document}